\documentclass[12pt,a4paper]{article}

\usepackage{amsfonts}

\usepackage{amssymb}

\usepackage[vcentermath,enableskew]{youngtab}
\def\e{\hbox{e}}
\def\Lap{{\cal L}}
\def\Dots{\cdot\cdot}
\newcommand{\CP}{\mathbb{C}\mathrm{P}}
\newcommand{\RP}{\mathbb{R}\mathrm{P}}
\newcommand{\cM}{{\cal{M}}}

\begin{document}
\title{\hfill\raise 20pt\hbox{\small DIAS-STP-03-5}\\
A Fuzzy Three Sphere and Fuzzy Tori}
\author{Brian P. Dolan$^{a),b)}$\footnote{\tt bdolan@thphys.may.ie}
\ and Denjoe O'Connor$^{b)}$\footnote{\tt denjoe@stp.dias.ie}\\
$^{a)}$Dept. of Mathematical Physics, NUI, Maynooth, Ireland\\
$^{b)}$School of Theoretical Physics, \\ Dublin Institute for Advanced Studies,\\ 
10~Burlington Rd., Dublin 8, Ireland\\
}

\maketitle

\begin{abstract}
A fuzzy circle and a fuzzy 3-sphere are constructed as subspaces of
fuzzy complex projective spaces, of complex dimension one and three,
by modifying the Laplacians on the latter so as to give unwanted
states large eigenvalues. This leaves only states corresponding to
fuzzy spheres in the low energy spectrum (this allows the commutative
algebra of functions on the continuous sphere to be approximated to
any required degree of accuracy). The construction of a fuzzy circle
opens the way to fuzzy tori of any dimension, thus circumventing the
problem of power law corrections in possible numerical simulations on
these spaces.
\end{abstract}

\section{Introduction}
\label{intro}

One of the principal goals of the study of field theories on fuzzy
spaces is to develop an alternative non-perturbative technique to the
familiar lattice one \cite{Montvay_Munster_book}. To date, this new
approach in the case of four dimensional field theories has been
limited to studies of Euclidean field theory on $S^2\times S^2$
\cite{SachinBadis}, $\CP^2$ \cite{GarnikBalGiorgioBadis} and 
$S^4$ \cite{DJ}. All but $S^2\times S^2$ have additional
complications.  For example, $\CP^2$ is not spin but spin{}$^c$ and
$S^4$ is really a squashed $\CP^3$ and includes many unwanted massive
Kaluza-Klein type modes. Even $S^2\times S^2$ is not ideal since it
has curvature effects that drop off as power corrections rather than
exponentially as in the case of toroidal geometries. 

The fuzzy approach does, however, have the advantage of preserving
continuous symmetries such as the $SU(2)$ symmetry of a round $S^2$
and does not suffer from fermion doubling \cite{FermionDoubling}. The
advantages are gained at the cost of introducing a non-locality
associated with the non-commutativity of the fuzzy sphere.  There is
therefore a balance of advantages and disadvantages associated with
the fuzzy approach. The final decision on whether the approach has
real advantages over the standard lattice approach should be
determined by doing genuine simulations.  For this reason Monte Carlo
simulations of the fuzzy approach are now in progress. In the lattice
approach non-locality is also a problem when fermions are included.
So our expectation is that as far as Monte Carlo simulations are
concerned the fuzzy approach will not be competitive with the lattice
one until fermions are included. The approach will gain further
advantages in situations where symmetries are more important. It also
extends naturally to allow for supersymmetry.  (see
\cite{bal_seckin_efrain} where a fuzzy supersphere was constructed).
So we expect the true power of the approach to emerge when
supersymmetry and chiral symmetry are present in a model.

A radically different alternative to the Euclidean Monte Carlo
approach becomes available once one has a fuzzy three-dimensional
space. Such a space has the advantage that it allows one to develop
very different non-perturbative methods, since now one can address the
non-perturbative questions from a Hamiltonian point of view.

The purpose of this article is to introduce precisely such fuzzy
three-dimensional spaces.  We will begin by presenting a fuzzy version
of the circle $S^1_F$, from which one can obtain tori of arbitrary
dimension. We will then present a fuzzy approximation to the
three-sphere, $S^3_F$.  Unfortunately, both of these spaces are still
not ideal in that they involve many unwanted additional degrees of
freedom which we suppress so that they do not contribute to the low
energy physics.  The presence of additional degrees of freedom is
probably unavoidable as it seems to be the price one pays for the
classical space not being a phase space.  The three-sphere is also
curved and hence the results obtained from studies of field theories
on this space will approach those of a flat three-dimensional space
with polynomial corrections.  It has, however, the advantages of
preserving the full $SO(4)$ symmetry of a round $S^3$. From the
construction it seems clear that both of these spaces will also be
free of fermion doubling problems.

We will restrict our focus here to scalar field theories and
demonstrate how the unwanted degrees of freedom can be suppressed so
that the limiting large matrix theory of a scalar field theory
recovers field theory on the commutative spaces. We will argue that
the data specifying the geometries can be cleanly specified by giving
a suitable Laplace-type operator for the scalar field, which together
with the matrix algebra and its Hilbert space structure gives a
spectral triple.

Aside from our personal motivations, non-commutative geometry has
recently become a very popular area of research from both the point of
view of possible new physics in string theory and $D$-brane theory,
\cite{ARS,HNS}, and as a new regularisation technique in ordinary
quantum field theory, \cite{SachinBadis}-\cite{DJ} and
\cite{GrosseKlimcikPresnajder}-\cite{Bal_etal}.  In both
these endeavours ``fuzzy'' spaces play an important r\^ole.  Roughly
speaking a fuzzy space is a finite matrix approximation to the algebra
of functions on a continuous manifold, the seminal example being the
fuzzy two-sphere, \cite{Madore}.  It has the important property of
preserving the isometries of the space that it is approximating.  As
such the idea can serve as a source of examples related to matrix
models in string theory and as a regularisation technique for ordinary
quantum field theory.  As a regularisation method it provides one that
preserves the underlying space-time symmetries and is amenable to
numerical computation.

Fuzzy spheres in dimensions other than two were analysed in
\cite{Ramgoolam}-\cite{Ho:Ramgoolam},
but the construction there was incomplete. They also advocate
projecting out the unwanted modes and working with a non-associative
algebra which we consider unsatisfactory. Also the case of odd spheres
works very differently to that of even spheres.  An alternative
approach for the fuzzy four-sphere, $S^4_F$, was given in \cite{DJ},
based on the fact that fuzzy $\CP^3$ and $\CP^1\cong S^2$ are well
understood \cite{CPN}, and, in the continuum limit, $\CP^3$ is an
$S^2$ bundle over $S^4$.  

In this paper we show how the odd-dimensional fuzzy spheres $S^1_F$
and $S^3_F$ can be extracted from the matrix algebras associated with
the fuzzy complex projective spaces $\CP^1_F$ and $\CP^3_F$.  An
alternative approach to obtaining a finite approximation to $S^3\cong
SU(2)$, based on conformal field theory, was presented in
\cite{FroehlichGawedzki}, however, in this approach it is unclear how
the unwanted modes are to be suppressed.  Our method uses a similar
suppression mechanism to that used for $S^4_F$ in \cite{DJ}. Although
there is no closed finite dimensional matrix algebra for $S^N_F$
unless $N=2$, the relevant degrees of freedom when $N=1$ and $N=3$ are
contained in the matrix algebras for $\CP^1$ and $\CP^3$ respectively.
One can therefore obtain functional integrals for field theories over
$S^1_F$ and $S^3_F$ by starting with functional integrals over
$\CP_F^1$ and $\CP^3_F$ and then suppressing the unwanted modes so
that they do not contribute to the functional integral.  Because of
the high degree of symmetry inherent in the construction, the unwanted
modes can be suppressed simply by using appropriate quadratic Casimirs
in the Laplacian. In this way we by-pass the problems associated with
the fact that the algebra of matrices associated with functions on the
sphere does not close on the sphere, but necessarily lifts into the
enveloping complex projective space.  In a similar fashion we expect
that when a Hamiltonian approach to field theory is developed using
these spaces the unwanted modes will cause no difficulties since they
can be made arbitrarily difficult to excite.

The paper is organized as follows. In section \ref{geometry} we
summarize how a given geometry is captured in the fuzzy
approach. Section \ref{circle} then gives our construction of a fuzzy
circle, $S^1_F$.  Section \ref{4sphere} summarises the construction of
$S^4_F$ presented in \cite{DJ} and in section \ref{3sphere} we present
our fuzzy three-sphere, $S^3_F$. Section \ref{3sphere:v2} gives an
alternative construction of $S^3_F$ which lends itself to a
generalisation to $S^N_F$ for any $N$ \cite{TSNF}.

\section{Encoding the geometry of a fuzzy space}
\label{geometry}

Fr\"ohlich and Gaw\c{e}dzki \cite{FroehlichGawedzki} (following
Connes, \cite{Connes:book}) have demonstrated that the
abstract triple $(H,{\cal{A}},\Delta_{\gamma})$, where
$H$ is the Hilbert space of square integrable functions on
the manifold $\cM$, with Laplace-Beltrami operator $\Delta_\gamma$,
$\gamma$ being the metric, and $\cal{A=C^{\infty}(\cM)}$ is the
algebra of smooth bounded functions on $\cM$, captures a topological
space together with its metrical geometry.

In a similar fashion one can specify a fuzzy space, $\cM_F$, as the
sequence of triples
\begin{equation}
\cM_F := (H_L,\mathrm{Mat}_{d_L},\Delta_L)
\label{fuzzy_triple}
\end{equation}
parameterized by
$L$, where $H_L=\mathbb{C}^{d_L}$ is the Hilbert space acted on
by the complete matrix algebra $\mathrm{Mat}_{d_L}$ of dimension $d_L$
with inner product $<M,N>={1\over{d_L}}Tr(M^\dag N)$ and $\Delta_L$ is a
suitable Laplacian acting on matrices. One can readily extract
information such as the dimension of the space from these data.  The
Laplacian comes with a cutoff and so the dimension can be read from
the growth of the number of eigenvalues. Sub-leading corrections give
such quantities as the Euler characteristic and 
other information about the space.

The data contained in the triple
$(H,{\cal{A}},\Delta_{\gamma})$ are precisely the data that
go into the Euclidean action for a scalar field theory on the space
$\cM$ and hence specifying the scalar action is a convenient
method of prescribing these data.

In the fuzzy approach the algebra will always be a matrix algebra and
we will retain the Hilbert space inner product specified above so the
only data from the triple, $(H_L,\mathrm{Mat}_{d_L},\Delta_L)$,
remaining to be supplied are the permitted matrix dimensions, $d_L$ and
a realization of the Laplacian, $\Delta_L$. Once this information is
given the fuzzy geometry is specified.

Though it may be convenient to give a map to functions this is not
necessary. Once the Laplacian is given its eigenmatrices and spectrum
can be used to provide such a map if needed. Suppose for example that
the spectrum of $\Delta_L$ is identical to that of $\Delta_{\gamma}$
up to some cutoff and a complete set of eigenmatrices is given by
$\hat\Psi_{\lambda}$ with the corresponding commutative eigenfunctions
being $\Psi_{\lambda}$, then the symmetric symbol-map $\mathrm{D}$
given by
\begin{equation}
\mathrm{D}=\sum_{\lambda}^{d_L^2}\Psi_{\lambda}\hat\Psi_{\lambda}
\label{symbol_map}
\end{equation}
provides a map to functions with
\begin{equation}
f_M={1\over d_L}Tr(\mathrm{D} M)
\end{equation}
the function corresponding to the matrix $M$. By construction the map
has no kernel and the symbol-map induces a $*$ product on functions
given by
\begin{equation}
f_M*_{\mathrm{D}}f_N={1\over d_L}Tr(\mathrm{D} MN)
\end{equation}
which represents matrix multiplication in terms of an operation on the
image functions. The $*$ product depends on $\mathrm{D}$, a different
but equivalent one could be obtained by giving a nonzero weighting
$c_{\lambda}(L)$ to the different terms in the sum
(\ref{symbol_map}). In the case of $\CP^{N}$ a particular choice of
the $c_{\lambda}(L)$ will give the diagonal coherent state
prescription\footnote{In the case where the symbol-map is the
projector of coherent states the function $f_M$ is referred to as the
covariant symbol of the matrix $M$ and since the coefficients
$c_{\lambda}(L)$ are not one it will differ from the corresponding
contravariant symbol, see Berezin \cite{Berezin}. The symbol-map is
referred to as symmetric when its covariant and contravariant symbols
are equal and coincides with the case of $c_{\lambda}(L)=1$.} as
discussed in \cite{CPN}.

If the symbol-map (\ref{symbol_map}) has the property that 
\begin{equation}
\Delta_{\gamma} f_M ={1\over d_L}Tr(\mathrm{D} \Delta_L M)
\end{equation}
where $\Delta_{\gamma}$ is a natural Laplacian for the space to be
approximated, then the spectrum of the fuzzy space will be precisely 
a cutoff version of that of the commutative space $\cM$. This is 
precisely what happens in the case of $\CP^N_F$, see \cite{CPN}.

However, it is convenient to extend the definition of fuzzy space to
the case where the spectrum coincides for low-lying eigenvalues, but
deviates for a family of eigenvalues that can be given arbitrarily
high value and which correspond to degrees of freedom that have no
counterpart in the commutative space $\cM$.  This allows us to obtain
fuzzy approximations to additional spaces --- in particular, as we will
see, to tori and the three sphere.

If one takes the Euclidean quantum field theory point of view
then the desired geometry appears as that associated with 
the accessible configurations of the field theory and the deviations
are suppressed in a probabilistic fashion.

A successful method of suppressing the unwanted modes would be to add
to the scalar action a term $S_I[\Phi]$ which is non-negative for any
$\Phi$, zero only for matrices that correspond to functions on
$\cM$, and positive for those that do not. The modified action would
therefore be of the form $S[\Phi]+h S_I[\Phi]$. The parameter $h$
should be chosen to be large and positive.  The probability of any
given matrix configuration then takes the form
\begin{equation}
{\mathcal P}[\Phi]=\frac{{\rm e}^{-S[\Phi]-hS_I[\Phi]}}{Z}
\label{prob_of_config}
\end{equation}
where 
\begin{equation}
Z=\int d[\Phi] {\rm e}^{-S[\Phi]-hS_I[\Phi]}
\label{partition_fn}
\end{equation}
is the partition function of the model.
If the prescription is to work for free field theories, then $S_I[\Phi]$
should be at most quadratic in $\Phi$. This can then be thought of as
a modification of the Laplacian in the triple (\ref{fuzzy_triple}).

Furthermore the problem of UV/IR mixing in scalar theories can be
removed by including a higher derivative operator in the quadratic
term of the field theory such that it renders all diagrams finite when
the matrix size is sent to infinity. With such a prescription since
each diagram has a limiting commutative value in the large matrix
limit each diagram must take this value and hence no UV/IR mixing can
occur.  The prescription of sending the matrix size to infinity and
sending the coefficient of the irrelevant higher derivative operator
to zero do not commute. This prescription of adding an irrelevant
operator to the action is simpler than the normal ordering
prescription proposed in \cite{Brian_Denjoe_Peter} and works for any
dimension.

From the above discussion it should be clear that the entire problem
of constructing a fuzzy approximation to a space is the problem of
giving a suitable prescription for the matrix Laplacian.

\section{Approximating a circle from a fuzzy sphere}
\label{circle}

Consider the finite matrix algebra representation of the fuzzy sphere
$S_F^2$ \cite{Madore}.  The algebra of $(L+1)\times (L+1)$ matrices, which
will be denoted by  $\mathrm{Mat}_{L+1}$, has the same dimension as the number
of degrees of freedom in a spherical harmonic expansion of a function
on $S^2$, truncated at angular momentum $L$,
\begin{equation}
f_L(\theta,\phi)=\sum_{l=0}^L \sum_{m=-l}^l f_{lm} Y_{lm}(\theta,\phi).
\label{YLm}
\end{equation}
That is\begin{equation}
\sum_{l=0}^L(2l+1)=(L+1)^2.
\end{equation}
The precise identification between a matrix $\Phi\in
\mathrm{Mat}_{L+1}$ and a cut-off function $f_L(\theta,\phi)$, as
discussed in the preceding section, is not unique, but the possible
maps can be given in terms of coherent states or the symmetric symbol-map $\mathrm{D}$ of (\ref{symbol_map}), and the resulting product of
functions is non-commutative for finite $L$.  It is crucial to our
construction that only maps for which the product of functions becomes
commutative in the limit $L\rightarrow\infty$ be considered.  The
symbol-map (\ref{symbol_map}) associates orthonormal $(L+1)\times
(L+1)$ polarisation tensors $\hat Y_{lm}$ with spherical harmonics
$Y_{lm}(\theta,\phi)$.  The conventions used here will be that
\begin{equation}
\hat Y_{lm}={1\over\sqrt{L+1}}\hat T_{lm}
\label{Tlm}
\end{equation}
where the polarisation tensors $\hat T_{lm}$ are those of \cite{VMK}.

The $SO(3)$ symmetric Laplacian, $\Lap^2$, on the fuzzy sphere acts on matrices
$\Phi$ and is represented by the second order Casimir corresponding to the
adjoint action of the angular momentum generators $L_i$ in the $(L+1)\times (L+1)$
representation:
\begin{equation}
\Lap^2\Phi=[L_i,[L_i,\Phi]].
\end{equation}
Hence the action can be taken to be
\begin{equation}
S[\Phi]={1\over L+1} Tr\left( {1\over 2}\Phi^\dagger \Lap^2\Phi +V(\Phi)\right)
\label{FS2action}
\end{equation}
for some scalar potential $V(\Phi)^\dagger=V(\Phi)$, which is assumed to be
bounded below. 
This action can then be used in a partition function which
involves ordinary integration over $(L+1)^2$ degrees of freedom
\begin{equation}
Z=\int{\cal D}\Phi\e^{-S[\Phi]}.
\label{Z}
\end{equation}
The probability distribution for field configurations is then
\begin{equation}
{\mathcal P}[\Phi]=\frac{{\rm e}^{-S[\Phi]}}{Z}
\label{prob_of_S2}
\end{equation}
where $S[\Phi]$ given is by (\ref{FS2action}). This probability
distribution is associated with the geometry
$(H_L,\mathrm{Mat}_{L+1},\Lap^2)$ which specifies a round fuzzy
sphere. The field theory with quadratic potential, however, suffers
from UV/IR mixing problems \cite{Brian_Denjoe_Peter,Chu_etal}. If we
add the term $a\Lap^4$ to the Laplacian and use the triple
$(H_L,\mathrm{Mat}_{L+1},\Lap^2+a\Lap^4)$ the UV/IR mixing problem is
removed and we recover a field theory on the commutative $S^2$ in the
infinite matrix size limit. The parameter $a$ can finally be sent to
zero with the result that the critical value of the mass parameter
will be sent to infinity. The process of taking the large matrix limit
and sending $a$ to zero do not commute. To obtain the commutative
theory on the sphere the matrix size must be sent to infinity for
non-zero $a$.

There is no finite matrix approximation to the algebra of functions on
$S^1$.  Nevertheless, the degrees of freedom relevant to a circle are
certainly contained in $\mathrm{Mat}_{L+1}$.  Focusing on the top
harmonic in (\ref{YLm}), with $l=L$, the $Y_{Lm}$ contain all $-L\le
m\le L$ and thus reproduce functions on the circle as
$m\rightarrow\infty$.  This implies that the partition function and
correlation functions for a field theory on a circle can be extracted
from that of the fuzzy sphere by suppressing all the modes with $l<L$
in (\ref{Z}).  One way of achieving this is to penalise modes with
$l<L$ by giving them a large positive weight in the action.  To this
end we modify the action (\ref{FS2action}) to
\begin{equation}
S_h[\Phi]={1\over L+1} Tr\left\{{1\over 2} \Phi^\dagger [L_3,[L_3,\Phi]]+
{h\over 2}\Phi^\dagger\left(L(L+1)-\Lap^2\right)\Phi
+V(\Phi)\right\}.
\end{equation}
All modes with $l<L$ now have the wrong sign for $\Lap^2$ and, when
$h$ is very large, are heavily penalised in the partition function
(\ref{Z}), contributing nothing as $h\rightarrow\infty$.  In this
limit only the modes with $l=L$ remain and these have the correct sign
for their kinetic energy, because the term linear in $h$ vanishes on
these and only these modes.  The `wrong sign' for the $\Lap^2$
contribution to the kinetic energy here is analogous to an
anti-ferromagnetic coupling in a lattice theory and just as in the
lattice theory with an anti-ferromagnetic coupling the action here is
also bounded below. That the action remains bounded from below is
intimately related to the fact that there is an ultraviolet cutoff in
the model and therefore a maximum eigenvalue for the Laplacian or
equivalently a shortest wavelength.

To see that the commutative algebra of functions on $S^1$ is recovered
in the $l=L$ sector of the fuzzy sphere as $L\rightarrow\infty$, we
first decompose the matrix $\Phi$ using the basis of polarisation
tensors:
\begin{equation}
\Phi=\sum_{l=0}^L \sum_{m=-l}^l \Phi_{lm}\hat Y_{lm}.
\end{equation}
In our conventions (\ref{Tlm}) the commutator of the polarisation
tensors is given by (see e.g. \cite{VMK} page 191, equation (46))
\begin{eqnarray}
[\hat Y_{l_1m_1},\hat Y_{l_2m_2}]&=&
\sqrt{(2l_1+1)(2l_2+1)\over L+1}
\sum_{l=0}^L(-1)^{L-l}\left\{1-(-1)^{l_1+l_2+l}\right\} \nonumber \\
&&\hspace{50pt}\times 
\left\{ \matrix{l_1 & l_2 & l \cr L/2 & L/2 & L/2\cr}\right\}
C_{l_1m_1,l_2m_2}^{lm}\hat Y_{lm},\nonumber\\
\end{eqnarray}
where $\left\{ \matrix{l_1 & l_2 & l \cr L/2 & L/2 & L/2\cr}\right\}$
are $6j$-symbols and $C_{l_1m_1,l_2m_2}^{lm}$ are Clebsch-Gordon
co-efficients.
Now for large $L$ 
\begin{equation}
\left\{ \matrix{l_1 & l_2 & l \cr L/2 & L/2 & L/2\cr}\right\}\approx
{1\over\sqrt{L+1}}C_{l_10,l_20}^{l0}
\end{equation}
and $C_{l_10,l_20}^{l0}=0$ when $l_1+l_2+l$ is odd.  Thus
\begin{equation}
[\hat Y_{l_1m_1},\hat Y_{l_2m_2}]\rightarrow 0
\end{equation}
and the algebra is commutative when $L\rightarrow 0$ as promised.  In
particular
\begin{equation}
[\hat Y_{Lm_1},\hat Y_{Lm_2}]\rightarrow 0
\end{equation}
and the top harmonic alone reproduces the commutative algebra
of functions on $S^1$ in the continuum.

To summarize we can encode the geometry specifying a fuzzy circle
by the triple 
\begin{equation}
S^1_F:= 
\left(H_L,\mathrm{Mat}_{L+1},\Lap^2_3+h\left(L(L+1)-\Lap^2\right)\right).
\label{S1Fdef}
\end{equation}
with $h>>1$.  This picks out the fuzzy circle from the top angular
momentum polarization tensor $\hat Y_{L,m}$. 

One could equally pick it out from a lower one, $\hat Y_{L_0,m}$ by
modifying the term proportional to $h$ to
$\left(L_0(L_0+1)-\Lap^2\right)^2$. This latter choice may have
advantages for the suppression of UV/IR mixing effects in the fuzzy
context. It roughly corresponds to a mixture of `nearest neighbour'
and next nearest neighbour ferromagnetic and anti-ferromagnetic
couplings.

Having constructed a fuzzy circle it is now clear that there is
no obstacle to constructing fuzzy tori of arbitrary dimension,
simply by taking products of fuzzy circles.  This has the
obvious advantage for numerical simulation of avoiding 
power-law curvature effects.

\section{Approximating $S^4$ from $\CP^3_F$}
\label{4sphere}

We can use a similar procedure to approximate $S^3$ from 
a finite approximation to $S^4$ but first, in this section,
we summarise the construction of 
the fuzzy $S^4$ from fuzzy $\CP^3$ given in \cite{DJ}.
The construction utilises the fact that $\CP^3$ is an $S^2$
bundle over $S^4$ and there is a well-defined matrix approximation
to $\CP^3\cong SU(4)/U(3)$.
The harmonic expansion of a function on $\CP^3$ requires representations
of $SU(4)$ that contain singlets of $U(3)$ under 
$SU(4)\rightarrow SU(3)\times U(1)$: in terms of $SU(4)$ Young tableaux
the permitted representations are
\begin{equation}
\overbrace{\young(\ \Dots \ ,\ \Dots \  ,\ \Dots \  )}^n
\kern -2.2pt \raise 13.4pt\hbox{$\overbrace{\young(\ \Dots \ )}^n$}
\label{HarmCP3}
\end{equation}
and are of dimension ${1\over 12}{(2n+3)(n+2)^2(n+1)^2}$.
All such representation, for $n\le L$, appear in the tensor 
product
\begin{equation}
\overbrace{\young(\ \Dots \ ,\ \Dots \  ,\ \Dots \  )}^L
\times\raise 13.4pt 
\hbox{$\overbrace{\young(\ \Dots \ )}^L$} ={\bf 1}+{\bf 15}+{\bf 84}
+\cdots + {(2L+3)(L+2)^2(L+1)^2\over 12}.
\label{MatCP2}
\end{equation}
Since the dimension of  
$\overbrace{\young(\ \Dots \ )}^L$ is $d_L:={1\over 6}{(L+3)(L+2)(L+1)}$
the representations in (\ref{MatCP2}) constitute a $d_L\times d_L$
matrix and are thus in one-to-one correspondence with elements $\Phi$
of $\mathrm{Mat}_{d_L}$.

Fuzzy $\CP^3$ is now identified with $\mathrm{Mat}_{d_L}$ with an appropriate
Laplacian.  The most natural Laplacian on $\CP^3_F$ is the $SU(4)$ invariant one
which is the the second order Casimir corresponding to the adjoint action
of the $SU(4)$ generators in the $d_L\times d_L$ representation.
For future convenience we shall use the fact that $SU(4)\approx Spin(6)$
and denote the generators by $J_{AB}$, with $A,B=1,\ldots, 6$ and 
$J_{AB}=-J_{BA}$.  The $Spin(6)$ invariant Laplacian on $\mathrm{Mat}_{d_L}$ is then
\begin{equation}
\Lap^2_{(6)}\Phi={1\over 2}[J_{AB},[J_{AB},\Phi]].
\label{SO6Lap}
\end{equation}
As $L\rightarrow \infty$ this corresponds to the continuum $Spin(6)$
invariant Laplacian on $\CP^3$.

In the notation of \cite{Fulton+Harris} 
we shall label the $Spin(6)$ irreducible representations by
their highest weights $(n_1,n_2,n_3)$, with $n_1,n_2,n_3$ either
all integers or all half-integers and $n_1\ge n_2 \ge n_3$.
The dimensions of these representations are
\begin{equation}
d^{(6)}=(n_1,n_2,n_3)={1\over 12}\Bigl((n_1+2)^2-(n_2+1)^2\Bigr)
\Bigl( (n_1+2)^2-n_3^2\Bigr)\Bigl((n_2+1)^2-n_3^2\Bigr)
\end{equation}
and the eigenvalues of $\Lap^2_{(6)}$ are
\begin{equation}
C^{(6)}_2(n_1,n_2,n_3)=\Bigl( n_1(n_1 +4) + n_2(n_2+2) +n_3^2\Bigr).
\end{equation}
The irreducible representations (\ref {HarmCP3}) that appear in a
harmonic expansion of function on $\CP^3$ are $(n,n,0)$ with $n$ an
integer, so the quadratic Casimir takes the value
\begin{equation}
C_2^{(6)}(n,n,0)=2n(n+3).
\label{SO6Casimir}
\end{equation}
In this notation (\ref{MatCP2}) translates to
\begin{equation}
\overline{\bigl({L\over 2},{L\over 2},{L\over 2}\bigr)}\times
\bigl({L\over 2},{L\over 2},{L\over 2}\bigr)=\sum_{n=0}^L(n,n,0).
\end{equation}

The extraction of a fuzzy $S^4$ from this algebra further relies on
the curious fact that there is another possibility for the
Laplacian on $\CP_F^3$ that has a lower symmetry, $SO(5)$,
coming from the fact that it is also
possible to represent $\CP^3$ as the coset
space $SO(5)/\Bigl(SU(2)\times U(1)\Bigr)$.
In this representation the harmonic expansion of a function on $\CP^3$
requires all representations of $Spin(5)$ that contain singlets
of $SU(2)\times U(1)$ under the decomposition
\begin{eqnarray}
Spin(5) &\rightarrow & SU(2)\times U(1)\nonumber\\
{\bf 4} & \rightarrow & 
   {\bf 2}_0 +{\bf 1}_1+{\bf 1}_{-1}\nonumber\\
\label{CP3embedding}
{\bf 5} & \rightarrow & {\bf 2}_1+{\bf 2}_{-1} +{\bf 1}_0\\
{\bf 10} & \rightarrow & {\bf 3}_0+{\bf 2}_1+{\bf 2}_{-1} +{\bf 1}_2
+{\bf 1}_{-2}+{\bf 1}_0.\nonumber
\end{eqnarray}
Irreducible representations of $Spin(5)$ can be labelled
by two numbers $(n_1,n_2)$, either both integers or both half-integers,
and $n_1\ge n_2$.  They have dimension
\begin{equation}
d^{(5)}(n_1,n_2)={1\over 6}(n_1-n_2+1)(n_1+n_2+2)(2n_1+3)(2n_2+1)
\end{equation}
and second order Casimirs\begin{equation}
C_2^{(5)}(n_1,n_2)=\Bigl( n_1(n_1+3) + n_2(n_2+1)\Bigr).
\label{SO5Casimir}
\end{equation}
From (\ref{CP3embedding}) we see that 
the $Spin(5)$ representations that contain singlets of $SU(2)\times U(1)$
are those with $(n_1,n_2)$ both integers --- these are
all the tensor representations $T_{a_1\cdots a_n}$, with $a_j=1,\ldots 5$,
and are therefore really representations of $SO(5)$.
The $SO(6)$ representations appearing in
(\ref{MatCP2}) decompose into $SO(5)$ representations as
\begin{eqnarray}
SO(6) & \rightarrow & SO(5)\nonumber\\
{\bf 15} & \rightarrow & {\bf 5} + {\bf 10}\\
{\bf 84} & \rightarrow & {\bf 14}+{\bf 35} +{\bf 35'},\nonumber\\
& \kern -278pt \hbox{or, in general,} &\nonumber\\
(n,n,0)& \rightarrow &\sum_{m=0}^n(n,m).
\label{SO5Mat}
\end{eqnarray}

The fact that $\CP^3\cong SO(5)/\Bigl(SU(2)\times U(1)\Bigr)$ means
that $SO(5)$ acts transitively on $\CP^3$ and functions on $\CP^3$ can
be expanded in terms of $SO(5)$ irreducible representations $(n,m)$
with an $SO(5)$ invariant Laplacian.  As discussed in \cite{DJ}, there
is no unique $SO(5)$ invariant Laplacian on $\mathrm{Mat}_{d_L}$ but
rather any linear combination of the restrictions of (\ref{SO6Lap}) to
$SO(5)$: {\it i.e.}~any linear combination of 
\begin{equation}
\Lap^2_{(5)}\Phi={1\over 2}[J_{ab},[J_{ab},\Phi]]
\end{equation}
and
\begin{equation}
\Lap^2_{(v)}\Phi=[J_{a},[J_{a},\Phi]],
\end{equation}
with $a,b=1,\ldots, 5$ and where $J_{a}=J_{a6}$, can be used as a
Laplacian provided the combination has positive eigenvalues.

The fuzzy $S^4$ can now be extracted from this by noting that the
harmonic expansion of functions on $S^4\cong SO(5)/SO(4)$ require
irreducible representations of $SO(5)$ that contain singlets of
$SO(4)$ under the restriction of $SO(5)$ to $SO(4)$. These are of
course the symmetric tensor representations of $SO(5)$, labelled by
$(n,0)$ in the notation above.  A Laplacian whose low lying modes are
those of $S_F^4$ can be constructed by penalizing the modes $(n,m)$ in
(\ref{SO5Mat}) with $m\ne 0$.  From (\ref{SO6Casimir}) and
(\ref{SO5Casimir}) we see that
\begin{equation}
2C_2^{(5)}(n,m)-C_2^{(6)}(n,n,0)=2m(m+1),
\end{equation}
so the Laplacian
\begin{equation}
\Lap^2_h = \Lap^2_{(6)}+h\Bigl(2\Lap^2_{(5)} -\Lap^2_{(6)}\Bigr)
\label{S4Lap}
\end{equation}
has eigenvalues $2n(n+3) + 2hm(m+1)$ and states with $m> 0$ will be
suppressed in a functional integral for large $h$.  The parameter $h$
here is acting like a ``squashing'' parameter, $h=0$ is the ``round''
$SO(6)$ invariant metric on $\CP^3$, while $h\neq0$ breaks this
symmetry down to $SO(5)$. The lowest permitted value for $h$ is
$h=-1$. We have $\Lap^2_{-1}=2\Lap^2_{v}$, and we see that this
Laplacian is rather singular in the large $L$ limit as the
representation $(n,n)$ for large $n$ develops a zero eigenvalue.

The family of actions
\begin{equation}
S_h[\Phi]={1\over d_L} Tr\left\{\Phi^\dagger \Lap^2_h \Phi +V(\Phi)\right\}
\label{FuzzyS4}
\end{equation}
gives a field theory on squashed $\CP_F^3$ for $h>-1$.  Furthermore as
$h\rightarrow\infty$ modes with $m>0$ are completely suppressed in a
functional integral and (\ref{FuzzyS4}) corresponds to a field theory
on $S_F^4$.  Note that it does not matter whether we use ${\cal
L}^2_{(6)}$, $2{\cal L}^2_{(5)}$ or $2{\cal L}^2_{(v)}$ for the first
term on the right-hand side of (\ref{S4Lap}) --- when the constraint
$m=0$ is imposed all three become the same operator.

\section{$S^3_F$ from $S^4_F$}
\label{3sphere}

We can build on the construction of the last section to get a
Laplacian whose low lying modes are those associated with a field on
$S^3_F$ by using the same trick as in section \ref{circle} to pick out
the top mode $n=L$ of the $S_F^4$.  As an irreducible representation
of $SO(5)$ this is the representation $(L,0)$ with dimension
\begin{equation}
d^{(5)}(L,0)={1\over 6}(2L+3)(L+2)(L+1).
\label{MatFS3}
\end{equation}
The harmonic expansion of a function on $S^3\cong SU(2)$ 
requires all irreducible representations of $SU(2)$, both integral
and half-integral,
\begin{equation}
f_L(\theta,\phi,\psi)=\sum_{j=0,{1\over 2},\ldots}^{L/2}\sum_{\overline m=-j}^j
\sum_{m=-j}^j f^j_{\overline m,m}D^j_{\overline m,m}(\theta,\phi,\psi),
\label{S3Harmonic}
\end{equation}
where $(\theta,\phi,\psi)$ are Euler angles and $D^j(\theta,\phi,\psi)$ 
are the Wigner $D$-matrices.
The key to extracting $S_F^3$ from $S_F^4$ is the observation 
that the 
total number of degrees of freedom in (\ref{S3Harmonic}) is
\begin{equation}
\sum_{j=0,{1\over 2},\ldots}^{L/2}(2j+1)^2={1\over 6}(2L+3)(L+2)(L+1)
\label{MatFS2}
\end{equation}
which is the same as $d^{(5)}(L,0)$ in (\ref{MatFS3}).  This is
because the top mode (or indeed any mode $(n,0)$) of $S_F^4$ has the
representation content of an $S_F^3$.

The top $SO(5)$ mode of $S_F^4$ can now be picked out by penalising
the modes with $n<L$ with an `anti-ferromagnetic' kinetic-energy term.
To this end we define the Laplacian
\begin{equation}
\Lap^2_{h,h'}\Phi={1\over 2} [J_{\alpha\beta},[J_{\alpha\beta},\Phi]] +
h'\Bigl( 2L(L+3)-{\cal L}^2_{(6)}\Bigr)\Phi+ h\Bigl(2\Lap^2_{(5)}-\Lap^2_{(6)}\Bigr)\Phi,
\label{FS3def}
\end{equation}
with $\alpha,\beta=1,\ldots,4$.  For finite $h$ and $h'$, as both the
last two terms are $>0$, this is a positive operator and contains all
modes on $\CP^3_F$.  As $h\rightarrow \infty$ modes not relevant to
$S^4_F$ are sent to infinity and, finally, modes not relevant to
$S_F^3$ are sent to infinity when $h'\rightarrow \infty$.  For very
large $h$ and $h'$, the low lying eigenvalues are therefore precisely
those of $S^3_F$. In a functional integral for a scalar field based on
this Laplacian we recover scalar field theory on $S_F^3$ in the large
$h$ and $h'$ limit.  Field theory on $S_F^3$ therefore arises from the
double limit $h,h'\rightarrow \infty$ in the action
\begin{equation}
S_{h,h'}[\phi]= {1\over d_L}Tr\left\{\Phi^\dagger \Lap^2_{h,h'}\Phi
+V(\Phi)\right\}
\end{equation}
with $\Phi\in \mathrm{Mat}_{d_L}$.

\section{An alternative construction of $S^3_F$}
\label{3sphere:v2}

The constructions described up till now have relied on matrix
approximations to $\CP^N$, specifically $\CP^1\cong S^2$ and $\CP^3$.
There is however another construction for $S_F^3$ based on
the orthogonal Grassmannian 
\hbox{$SO(5)/\Bigl(SO(3)\times SO(2)\Bigr)$}. This is a co-adjoint orbit and therefore a well-defined
finite matrix approximation to the algebra of functions on this
Grassmannian exists.  This space is not the same as $\CP^3$: it arises
from a different embedding of $SU(2)\times U(1)$ into $Spin(5)$,
characterised by the decomposition
\begin{eqnarray}
Spin(5) & \rightarrow & SU(2)\times U(1)\nonumber\\
{\bf 4} & \rightarrow & {\bf 2}_1 + {\bf 2}_{-1}\nonumber\\ 
{\bf 5} & \rightarrow & {\bf 3}_0 +{\bf 1}_2 + {\bf 1}_{-2}\\ 
{\bf 10} & \rightarrow & {\bf 3}_2+{\bf 3}_{-2} +{\bf 3}_0+{\bf 1}_0.\nonumber
\end{eqnarray}
In particular the ${\bf 5}$ does not give a singlet under this embedding
and so must be excluded from the harmonic expansion on this space ---
it is clearly not the same space as $\CP^3$.  The representation
content here is such that
\begin{equation}
SO(5)/\Bigl(SO(3)\times SO(2)\Bigr)\cong 
Sp(2)/U(2),
\end{equation}
which is known not to admit a spin structure \cite{DN}.  The expansion
of a function on the orthogonal Grassmannian $SO(5)/\Bigl(SO(3)\times
SO(2)\Bigr)$ can be obtained from that on $\CP^3\cong
SO(5)/\Bigl(SU(2)\times U(1)\Bigr)$ simply by omitting all the odd
rank tensors from the latter.  In this way $SO(5)/\Bigl(SO(3)\times
SO(2)\Bigr)$ follows from moding out the $SO(5)$ representation of
$\CP^3$ by the ${\mathbb Z}_2$ action $T_a-\rightarrow -T_a$ on the
${\bf 5}$, so
\begin{equation}
SO(5)/\Bigl(SO(3)\times SO(2)\Bigr)\cong \CP^3/{\mathbb Z}_2.
\end{equation}
As a side remark we note that 
$SO(5)/\Bigl(SO(3)\times SO(2)\Bigr)$ is an $S^2$ bundle
over the real projective space $\RP^4$. 

In the notation of section \ref{4sphere} the even rank tensor
representations of $SO(5)$ are $(n,m)$ with $n+m=2l$, and these have
dimension
\begin{equation}
d^{(5)}(2l-m,m)={1\over 6}(2l-2m+1)(2l+2)(4l-2m+3)(2m+1).
\end{equation}
So the total number of degrees of freedom for $m\le l$ and
$0\le l \le L/2$ ($L$ even) is
\begin{equation}
\sum_{l=0}^{L/2}\sum_{m=0}^{l}d^{(5)}(2l-m,m)
=\left[ {(L+4)(L+3)(L+2)\over 24}\right]^2,
\end{equation}
which is the same as that of $\mathrm{Mat}_{d'_L}$ with 
\begin{equation}
d'_L={1\over 24}{(L+4)(L+3)(L+2)}.
\label{dlprime}
\end{equation}
Thus for matrix dimensions $d'_L$ we have fuzzy orthogonal Grassmanians.
The action on this fuzzy Grassmannian looks essentially identical to that
on a squashed $\CP^3_F$, 
\begin{equation}
S[\Phi]={1\over d'_L}Tr\left\{  \Phi^\dagger \Lap^2_{(5)}\Phi + V(\Phi)
\right\},
\end{equation}
except that the matrix algebras are restricted to those of size $d'_L$
in (\ref{dlprime}) containing only even rank tensor representations
$SO(5)$, $d^{(5)}(2l-m,m)$ and one has only one quadratic Casimir, the
$SO(5)$ one, at ones disposal.

The harmonic expansion of a function on $S_F^3$ is contained in
$\mathrm{Mat}_{d_L'}$ because the top representation 
for a given $L$, $2l=L$ with $m=0$,
has dimension (\ref{MatFS3})
and, as observed in the previous section, this is a sum of the 
dimensions of the $SU(2)$ representations
required for a harmonic expansion on $S^3$, (\ref{MatFS2}).

If we can penalise all modes with $2l<L$ and $m\ne 0$ for $2l=L$
in a functional integral over $SO(5)/\Bigl(SO(3)\times SO(2)\Bigr)$
then we will really be doing a functional integral over $S_F^3$.
This is easily achieved since $2l=L$ and $m=0$
has the largest second order Casimir,
\begin{equation}
C_2^{(5)}(L,0)=L(L+3),
\end{equation}
of all the $SO(5)$ representations in $\mathrm{Mat}_{d_L'}$.
In the now familiar manner the unwanted modes in the
functional integral over $SO(5)/\Bigl(SO(3)\times SO(2)\Bigr)$
can be suppressed by using the
Laplacian
\begin{equation}
\Lap^2_{h'}=
{1\over 2}[J_{\alpha\beta},[J_{\alpha\beta},\cdot]] 
+h'\Bigl(L(L+3)
-\Lap^2_{(5)}\Bigr),  
\end{equation}
which acts on fields $\Phi\in \mathrm{Mat}_{d_L'}$ and $L$ even.
The unwanted modes are completely eliminated in the
limit $h'\rightarrow \infty$, giving $S_F^3$
truncated at level $L$.
The constraint  that $L$ is even does not change the
fact that we get the full continuum $S^3$ as $L\rightarrow\infty$.

\section{Conclusions}
\label{conclusions}

By starting with the known finite matrix algebras for $\CP^3$ and
$\CP^1$, the fuzzy $\CP^3_F$ and the fuzzy sphere $\CP^1_F\cong
S^2_F$, finite functional integrals for scalar field theories on
$S^3_F$ and $S^1_F$ have been constructed.  
The geometry of a fuzzy space is specified by a triple 
$(H_L,\mathrm{Mat}_{d_L},\Delta_L)$ and,
although there is no known closed associated algebra giving
a fuzzy $S^1$ as a triple directly,
$\CP^1_F$ nevertheless contains
the states required for a $S_F^1$ plus other unwanted states.
The unwanted states are given large eigenvalues by modifying
the Laplacian on $\CP^1_F$, as in equation (\ref{S1Fdef}),
leaving only the states of $S_F^1$ in the low energy
spectrum of the Laplacian.  
In a similar way $\CP_F^3$ contains the states necessary
for a fuzzy description of $S^3$ (via $S_F^4$) and the Laplacian on $\CP^3_F$
can be modified, as in equation (\ref{FS3def}), 
so that states not related to $S_F^3$ are given
large eigenvalues, leaving only $S_F^3$ states in the low energy
spectrum.

An alternative construction of $S_F^3$, based on suppressing modes
on a fuzzy version of the orthogonal Grassmannian
$SO(5)/\Bigl(SO(3)\times SO(2)\Bigr)$,
has been presented in section \ref{3sphere:v2}.  
This has the advantage of having a natural extension
to $S_F^N$ for any $N$, \cite{TSNF}.
Thus $S^3_F$ can be obtained
either in two steps, via the fuzzy $S^4_F$ constructed in \cite{DJ},
$\CP^3_F\rightarrow S^4_F \rightarrow S^3_F$, or alternatively in a
single step from the fuzzy version of $SO(5)/[SO(3)\times SO(2)]$ as
described in section \ref{3sphere:v2}.  

The construction of the fuzzy circle allows fuzzy tori to
be defined in an obvious way, by taking products of fuzzy circles,
thus opening the way to numerical simulations on tori while
preserving the full $U(1)\times \cdots \times U(1)$ isometry group
and avoiding the fermion doubling problem \cite{FermionDoubling}.
One first writes
down a finite functional integral for a field theory on
$S^2_F\times \cdots \times S^2_F$, which contain the modes relevant for 
propagation $S_F^1\times\cdots\times S_F^1$ in its spectrum, 
and then damps the unwanted modes.  This
can be done, in a manner that preserves the isometries of the torus,
by introducing appropriate combinations of second order Casimirs into
the propagators.  

It is pleasure to acknowledge A. P. Balachandran, Peter
Pre\v{s}najder, Harald Grosse and Daniel Roggenkamp for helpful
discussions.


\begin{thebibliography}{03}

 \bibitem{Montvay_Munster_book}       
I.~Montvay and G.~Munster, 
{\sl Quantum Field on a Lattice}, Cambridge, 1997. 

\bibitem{SachinBadis}
S.~Vaidya and B.~Ydri,
{\sl On the Origin of the UV-IR Mixing in Noncommutative Matrix Geometry},
{\tt hep-th/0305201}.

\bibitem{GarnikBalGiorgioBadis}
G.~Alexanian, A.P.~Balachandran, G.~Immirzi and B.~Ydri,
{\it J. of Geom. and Phys.} {\bf 42} (2002) 28,
{\tt hep-th/0103023}.

\bibitem{DJ} Denjoe O'Connor and Julieta Medina, 
{\sl Scalar Field Theory on Fuzzy $S^4$}, {\tt hep-th/0212170}.

\bibitem{FermionDoubling} 
A.P.~Balachandran, T.R.~Govindarajan and B.~Ydri,
{\it Mod.~Phys.~Lett.} {\bf A15} (2000) 1279, {\tt hep-th/9911087};
A.P.~Balachandran, T.R.~Govindarajan and  B.~Ydri,
{\sl Fermion doubling problem and noncommutative geometry II},
{\tt hep-th/0006216};
A.P.~Balachandran and G.~Immirzi,
{\sl The Fuzzy Ginsparg-Wilson Algebra: 
A Solution of the Fermion Doubling Problem},
{\tt hep-th/0301242}.

\bibitem{bal_seckin_efrain}
A.P.~Balachandran, S.~Kurkcuoglu and E.~Rojas,
{\it JHEP} {\bf 0207} (2002) 056,
{\tt hep-th/0204170}.


\bibitem{ARS} A.~Yu.~Alekseev, A.~Recknagel and V.~Schomerus,
{\it JHEP} {\bf 9909}, (1999) 023, {\tt hep-th/9908040}.

\bibitem{HNS} Y.~Hikida, M.~Nozaki and Y.~Sugawara,
{\it Nucl.\ Phys.} {\bf B617}, (2001) 117, 
{\tt hep-th/0101211}.

\bibitem{GrosseKlimcikPresnajder}
H.~Grosse, C.~Klim\v{c}\'{\i}k and P.~Pre\v{s}najder,
{\it Int.\ J.\ Theor.\ Phys.\ }{\bf 35}, (1996) 231,
{\tt hep-th/9505175}; 
H.~Grosse and A.~Strohmaier,
{\it Lett. Math. Phys.} {\bf 48}, (1999) 163,
{\tt hep-th/9902138}.

\bibitem{grklpra}
H.~Grosse, C.~Klim\v{c}\'{\i}k and P.~Pre\v{s}najder,
{\it Comm. Math. Phys.} {\bf 178}, (1996) 507;
H.~Grosse and P.~Pre\v{s}najder,
{\it Lett. Math. Phys.} {\bf 46}, (1998) 61.

\bibitem{Peter} P.~Pre\v{s}najder,
{\it J. Math. Phys.} {\bf 41} (2000) 2789, {\tt hep-th/9912050}.

\bibitem{Bal_etal}
A.P.~Balachandran and S.~Vaidya,
{\it Int. J. Mod. Phys.} {\bf A16}, (2001) 17, {\tt hep-th/9910129};
S.~Baez, A.P.~Balachandran, S.~Vaidya and B.~Ydri,
{\it Comm. Math. Phys.} {\bf 208}, (2000) 787, {\tt hep-th/9811169}.

\bibitem{Madore} J.~Madore, {\it Class.~Quant.~Grav.} {\bf 9} (1992) 69.


\bibitem{Ramgoolam} S.~Ramgoolam, 
{\it Nucl.~Phys.} {\bf B610} (2001) 461,
{\tt hep-th/0105006}.

\bibitem{Kimura} Yusuke Kimura, 
{\sl On higher dimensional fuzzy spherical branes}, {\tt hep-th/0301055}.

\bibitem{Ramgoolam:odd_spheres}
S.~Ramgoolam, 
{\it JHEP} {\bf 0210} (2002) 064,
{\tt hep-th/0207111}.

\bibitem{Ho:Ramgoolam} P.M.~Ho and S.~Ramgoolam,
{\it Nucl.Phys.} {\bf B627} (2002) 266,\hfill\break
{\tt hep-th/0111278}.

\bibitem{CPN} A.P.~Balachandran, B.P.~Dolan, J.~Lee, X.~Martin and D.~O'Connor,
{\it J.~Geom.~and~Phys.} {\bf 43} (2002) 184, {\tt hep-th/0107099}.

\bibitem{FroehlichGawedzki}
J.~Frohlich and  K.~Gaw\c{e}dzki,
{\sl Conformal Field Theory and Geometry of Strings},
Lectures given at Mathematical Quantum Theory Conference, 
Vancouver, Canada, 4-8 Aug 1993. 
Published in Vancouver 1993, 
Proceedings, Mathematical quantum theory, {\bf Vol. 1} 57-97,
{\tt hep-th/9310187}.

\bibitem{TSNF} B.P.~Dolan, Denjoe O'Connor and P.~Pre\v{s}najder, 
in preparation.

\bibitem{Connes:book}
A.~Connes.
{\it Noncommutative Geometry},
Academic Press, London, 1994.

\bibitem{Berezin} F.A.~Berezin, 
{\it Comm. Math. Phys.} {\bf 40}, (1975) 153.

\bibitem{Brian_Denjoe_Peter}  
B.P.~Dolan, Denjoe O'Connor and P.~Pre\v{s}najder,
{\it JHEP} {\bf 0203} (2002) 013, 
{\tt hep-th/0109084}.

\bibitem{VMK} D.A.~Varshalovich, A.N.~Moskalev and V.K.~Khersonskii,
{\it Quantum Theory of Angular Momentum}, World Scientific (1988)

\bibitem{Chu_etal} Chong-Sun Chu, J.~Madore and H.~Steinacker,
{\it JHEP} {\bf 0108} (2001) 038,
\hspace{0.2cm}  {\tt hep-th/0106205}.

\bibitem{Fulton+Harris} W.~Fulton and J.~Harris , 
{\it Representation Theory: a first course}, 
Graduate Texts in Mathematics 129, Springer, (1991).

\bibitem{DN} B.P.~Dolan and C.~Nash, 
{\it JHEP} {\bf 0207} (2002) 057, {\tt hep-th/0207007}.

\end{thebibliography}
\end{document}